\documentclass[aps,pra,preprintnumbers,nofootinbib,onecolumn]{revtex4}
\usepackage{bm}
\usepackage{latexsym}
\usepackage{dcolumn}
\usepackage{amsmath,amsfonts,amssymb}
\usepackage{graphicx,epsfig}
\usepackage{psfrag}
\usepackage{amsthm}

\def\be {\begin{equation}}
\def\ee {\end{equation}}
\def\bea {\begin{eqnarray}}
\def\eea {\end{eqnarray}}
\def\bc {\begin{center}}
\def\ec {\end{center}}
\def\bfg {\begin{figure}}
\def\efg {\end{figure}}
\def\bi {\begin{itemize}}
\def\ei {\end{itemize}}
\def\nn {\nonumber}

\def\la {\label}
\def\le {\left}
\def\ri {\right}

\def\a  {\alpha}
\def\b  {\beta}

\def\d  {\delta}

\def\s {\sigma}

\def\beq{\begin{equation}}
\def\eeq{\end{equation}}
\def\br{\begin{eqnarray}}
\def\er{\end{eqnarray}}
\newcommand{\eel}[1] {\label{#1}\end{equation}}

\newcommand{\bdm}{\begin{displaymath}}
\newcommand{\edm}{\end{displaymath}}

\begin{document}
%\preprint{gr-qc/0704.xxxx}
%\hspace{15cm} 03/31/2009\\
\title{Relativistic particle in a three-dimensional box}

\author{Pedro Alberto $^1$}\email[Email: ]{pedro@fis.uc.pt}
\author{Saurya Das $^2$} \email[Email: ]{saurya.das@uleth.ca}
\author{Elias C. Vagenas $^3$} \email[Email: ]{evagenas@academyofathens.gr}
\affiliation{$^1$Departamento de F\'{\i}sica and Centro de
F\'{\i}sica Computacional, Universidade de Coimbra, P-3004-516
Coimbra, Portugal}

\affiliation{$^2$Theoretical Physics Group, Dept of Physics and
Astronomy, University of Lethbridge, 4401 University Drive,
Lethbridge, Alberta, Canada T1K 3M4}

\affiliation{$^3$Research Center for Astronomy and Applied
Mathematics, Academy of Athens, \\
Soranou Efessiou 4,
GR-11527, Athens, Greece
}

\begin{abstract}
We generalize the work of Alberto, Fiolhais and Gil and solve
the problem of a Dirac particle confined in a 3-dimensional box.
The non-relativistic and ultra-relativistic limits are considered
and it is shown that the size of the box determines how
relativistic the low-lying states are. The consequences
for the density of states of a relativistic fermion gas are
briefly discussed.
\end{abstract}
%\pacs{123xxx}

\maketitle

%%%%%%%%%%%%%%%%%%%%%%%%%%%%%%%%%%%%%%%%%%%%%%%%%%%%%%%
%%%%%%%%%%%%%%%%%%%%%%%%%%%%%%%%%%%%%%%%%%%%%%%%%%%%%%%
%
%
%
%
%%%%%%%%%%%%%%%%%%%%%%%%%%%%%%%%
% Dirac equation
%%%%%%%%%%%%%%%%%%%%%%%%%%%%%%%

\section{Introduction}

The problem of a particle confined in an one-dimensional infinite square well
potential lies at the heart of non-relativistic quantum mechanics, being the simplest problem that
illustrates how the wave nature of bound particles implies that their energy is
quantized. The generalization of this problem to three dimensions is used for the statistical description
of a fermion gas and is the starting point for some many-body theories of fermions like the Thomas-Fermi model
of the atom.

The relativistic formulation and solution of the problem of a
spin-1/2 fermion with mass $m$ confined in a one-dimensional
square well potential was done by Alberto, Fiolhais and Gil
\cite{afg}. In this paper we review their approach and generalize
it to a 3-dimensional square well potential. The problem of a relativistic
particle confined in infinite square well potential is traditionally not
dealt with in the textbooks of Relativistic Quantum Mechanics, even in the
most comprehensive ones such as the one by Greiner \cite{Greiner}.

Recently, in the context of
quantum gravity phenomenology, applications of this result were presented.
In particular, by applying the Generalized Uncertainty Principle (GUP)
to a particle confined in a three-dimensional box, it is shown that
the length of the box must be quantized in terms of a fundamental
length, e.g. Planck length, and this indicates that the nature of space
may be fundamentally grainy \cite{Das:2010zf}.

In the present analysis, we choose an ansatz for the spinor inside the box which is
consistent with the non-relativistic problem and derive a
transcendental equation for the wave numbers allowed. We obtain
the non-relativistic and ultra-relativistic limits of this
equation and its solutions, showing also how they are related to
the ratio between the size of the containing box and the Compton
wavelength $\hbar/(mc)$ of the fermion. Finally, we discuss
briefly the differences regarding the density of states between
the usual non-relativistic fermion gas and the relativistic
fermion gas.

\section{Solution of Dirac Equation with a one dimensional infinite square well}
\par\noindent
The Dirac equation is written as \footnote{In this
section, we follow the formulation of \cite{afg}.}
\bea
H \psi &=& \le (c\, \vec \a \cdot \vec p + \b mc^2 \ri) \psi (\vec r)  \nn \\
&=& E\psi (\vec r)
\la{ham1}
\eea
where $\alpha_i~(i=1,2,3)$ and $\b$ are the Dirac matrices, for which we
use the following representation
\bea
\a_i =
\left( \begin{array}{cc}
0 & \sigma_i \\
\sigma_i & 0 \end{array} \right)~,~
\b =
\left( \begin{array}{cc}
I & 0 \\
0 & -I \end{array} \right)
\eea
where $\sigma_i$ are the Pauli matrices and $I$ is the 2-dimensional unit matrix.
It is evident that in $1$-spatial dimension and in the position representation,
say $z$, the Dirac equation is given by
\bea \le( -i\hbar c \a_z \frac{d}{dz} + \b mc^2 \ri)\psi(z) =
E\psi (z)~. \la{diraceqn1} \eea
The positive energy solutions read
\be
\psi = N ~e^{ikz} \left( \begin{array}{c}
\chi \\
{} \\
%\frac{\sigma_z \hbar k_0 c}{E+mc^2}
r \sigma_z \chi \end{array} \right) \\
{} \nn \\
\ee
where $m$ is the mass of the Dirac particle, $k$
is the wavenumber that satisfies the usual dispersion relation $E^2 = (\hbar k c)^2 + (mc^2)^2$,
$r \equiv \frac{\hbar k c}{E+mc^2}$ and
$\chi^\dagger \chi=1$.
Note that $r$ runs from $0$ (non-relativistic) to $1$ (ultra-relativistic),
$k$ could be positive (right moving) or negative (left moving), while
$N$ is a suitable normalization constant.
\par\noindent
As noted in \cite{afg}, to confine a relativistic particle in a box of length $L$
in a consistent way avoiding the Klein paradox (in which an increasing number of
negative energy particles are excited), one may take its
mass to be $z$-dependent as was done in
the MIT bag model of quark confinement \cite{bhaduri,thomas}
\bea
m(z) &=& M,~z<0~~~\mbox{(Region I)} \\
     &=& m,~ 0 \leq z \leq L~~~\mbox{ (Region II)} \\
     &=& M,~z>L~~~\mbox{ (Region III)}\ ,
\eea
where $m$ and $M$ are constants, and taking eventually the limit $M \rightarrow \infty$.
%This guarantees that there is no mixing with negative energy solutions.
This guarantees that the confinement process avoids the excitation of negative energy Dirac sea
particles, since
the plane wave energy spectrum of positive and negative energy solutions is separated by
at least twice their mass times $c^2$.
Thus, the general form of the wavefunction for a bounded Dirac particle
in a one dimensional box, vanishing when $|z|\to \infty$, can be written (in all three regions)
as
\bea
\psi_I &=& A~e^{-iKz}
\left( \begin{array}{c}
\chi \\
{} \\
%\frac{\sigma_z \hbar K_0 c}{E+Mc^2}
-R \sigma_z \chi
\end{array} \right)
\la{psi1}\\
%
%+ A_2~e^{-iKz}
%\left( \begin{array}{c}
%\chi \\
%{} \\
%\frac{-\sigma_z \hbar K_0 c}{E+Mc^2}
%-R \sigma_z \chi
%\end{array} \right)
%\nn \\
%
% PSI II
%
\psi_{II} &=& B~e^{ikz}
\left( \begin{array}{c}
\chi \\
{} \\
%\frac{\sigma_z \hbar k_0 c}{E+mc^2}
r\sigma_z \chi
\end{array} \right)
+ C~e^{-ikz}
\left( \begin{array}{c}
\chi \\
{} \\
%\frac{-\sigma_z \hbar k_0 c}{E+mc^2}
-r\sigma_z \chi
\end{array}
\right)
\label{psi2}\\
%
% PSI III
%
\psi_{III} &=& D~e^{iKz}
\left( \begin{array}{c}
\chi \\
{} \\
%\frac{\sigma_z \hbar K_0 c}{E+Mc^2}
R \sigma_z \chi
\end{array} \right)
%
%+ D_2~e^{-iKz}
%\left( \begin{array}{c}
%\chi \\
%{} \\
%%\frac{-\sigma_z \hbar K_0 c}{E+Mc^2}
%-R \sigma_z \chi
%\end{array} \right)
%\nn \\
\la{psi3}
\eea
where $K=\sqrt{E^2-(Mc^2)^2}/(\hbar c)$ (an imaginary wavenumber when $M> E/c^2$)
and $R=\hbar K c/(E+Mc^2)$.
Thus, in the limit $M\rightarrow \infty$, $K \rightarrow +i\infty$,
the terms associated with $A$ and $D$ go to zero.
%
%
%In addition, without loss of generality we choose $B=1$ and $C=e^{i\delta}$ where $\delta$ is
%a real number. It can be shown that if one chooses $|C| \neq 1$ then the energy
%of the relativistic particle is complex.
%
%
%
Now, boundary conditions akin to those for the Schr\"odinger
equation, namely $\psi_{II}=0$ at $z=0$ and $z=L$ will require
$\psi_{II}$ to vanish identically. Thus, they are disallowed. This
is related to the fact that the usual boundary conditions for
non-relativistic quantum mechanics cannot always be applied to
relativistic problems, as was shown by V. Alonso et al.
\cite{VAlonso}. Instead, we require the outward component of the
Dirac current to be zero at the boundaries (the MIT bag model).
This ensures that the particle is indeed confined within the box
\cite{thomas}.

\par\noindent
The conserved current corresponding to Eq.(\ref{diraceqn1}) can be shown to be
\bea
\la{j_z}
J_z = \bar\psi \gamma^z \psi ~,
\eea
where $\gamma^z\equiv\gamma^3$ is the $z$-component of the four-vector set of $4\times4$ gamma matrices $\gamma^\mu=(\gamma^0,\vec\gamma)$.
These matrices are related to the $\alpha_i$ and $\beta$ matrices
introduced earlier by $\gamma^0=\beta$ and
$\gamma^i=\beta\alpha_i$. The vanishing of the outward component of the Dirac current
$J^\mu = \bar\psi\gamma^\mu\psi$ at a
boundary is obtained by requiring that the condition $i\gamma\cdot n \psi = \psi$ holds there,
where $n$ is the unit four-vector normal to the space-time boundary \cite{thomas}, such that $n_\mu n^\mu=-1$.
For a static boundary $n^\mu=(0,\hat x)$, where $\hat x$ is the outward normal to the boundary surface.
Applying this to the wavefunction $\psi_{II}$ at $z=0$ and $z=L$ gives
\cite{afg}
\bea
i\b\a_z\psi_{II}\big|_{z=0} &=& \psi_{II}\big|_{z=0}  \la{mitbc1} \\
\mbox{and}~
-i\b\a_z\psi_{II}\big|_{z=L} &=& \psi_{II}\big|_{z=L} \la{mitbc2}
\eea
respectively. Using the expression for $\psi_{II}$ from (\ref{psi2}), we get
from (\ref{mitbc1}) and (\ref{mitbc2}) respectively
\bea
\la{cond1}
\frac{B+C}{B-C} &=& ir \\
\la{cond2}
\frac{B e^{ikL} + Ce^{-ikL}}{Be^{ikL}-Ce^{-ikL}}
&=& -ir ~,
\eea
which in turn yield
\bea
(i r -1)  = \frac{C}{B}(ir+1)  \hspace{1ex}
\la{cond3}&& \\
(ir +1)= \frac{C}{B}(ir-1) e^{-i2kL} ~.&&\hspace{1ex}
\la{cond4}
\eea
By eliminating $C/B$ between (\ref{cond3}) and (\ref{cond4}) one gets
\bea
\frac{i r -1}{ir +1}=e^{ikL}
\la{kl}
\eea
so that we finally arrive to  the transcendental equation
\be
\tan(k L)=\frac{2 r}{r^{2}-1}=-\frac {\hbar k}{mc}~.
\label{1dwavenumber}
\ee
The discrete solutions of this equation for the values of the wavenumber $k$ give the quantized energy levels
for a relativistic particle in the 1-dimensional box. Note that the solution with $k=0$
is excluded because, from (\ref{cond3}), $C=-B$ and the upper component would be proportional
to $\sin(kx)$ and therefore be identically zero inside the box. It is worth mentioning at this point
that Eq.~(\ref{cond3}) implies that $|B|=|C|$ and this condition could have as well be obtained by just requiring
that $J_z$, given by (\ref{j_z}), vanishes at the boundary, as remarked by Menon and Belyi \cite{menon}.
Note that this condition does not imply that $B=\pm C$, since at least one of $B$ or $C$ is not a real number
in this case, as can also be seen from Eq.~(\ref{cond3}) and the fact that $r\not=0$. Also one may note that
in order to obtain the eigenvalue equation (\ref{1dwavenumber}), one needs to know the ratio of the complex coefficients
$C$ and $B$, $C/B$, which we got from the MIT boundary condition,
and not only the ratio of their moduli.

To conclude this section, we comment briefly on the negative energy solutions of the
Dirac equation (\ref{diraceqn1}) in a one-dimensional box. These solutions can be found
from the negative energy solutions of the free Dirac equation travelling in opposite directions
in the $z$ axis. The new boundary condition is obtained from the corresponding condition for positive energy by applying
the charge conjugation operator. The resulting equation for the wavenumber $k$ is identical to
(\ref{1dwavenumber}) and therefore its discrete solutions $k_n$ are the same. This means that one has an overall
symmetric energy spectrum, in which for every discrete positive level with energy $E=E_n=\sqrt{\hbar^2k_n^2+m^2c^4}$ there is one with $E=-E_n$.
This could be expected, since in our case one has a Dirac Hamiltonian with a confining Lorentz scalar potential
which anti-commutes with the charge conjugation operator, as does the free Dirac Hamiltonian.
Since the Hamiltonian with the present boundary conditions
is Hermitian (see \cite{menon}), all wavefunctions belonging to distinct eigenvalues must be orthogonal to each other.
Thus the positive and negative energy solutions do not mix. Similar conclusions hold for a two and three dimensional confining box as well.
%
%
%
%%%%%%%%%%%%%%%%%%%%%%%%%%%%%%%%%%%%%%%%%%%%%%%%%%%%%
\section{Solution of Dirac Equation with a three-dimensional infinite square well}
%%%%%%%%%%%%%%%%%%%%%%%%%%%%%%%%%%%%%%%%%%%%%%%%%%%%%
%
%

The generalization of the problem of the previous section involves solving
the $3$-dimensional Dirac equation with a position-dependent mass of the form
\begin{equation}
m(\vec r)=\left\{
\begin{array}{cc}
  m &  \vec r\in V\\[2mm]
  M &  \vec r\not\in V
\end{array}\right.\ ,
\label{3d_pot}
\end{equation}
where $V$ is defined by the set of points with coordinates $(x_1,x_2,x_3)$ such that
$0\leq x_1\leq L_1,\,0\leq x_3\leq L_2,\,0\leq x_3\leq L_3 $.
As before we take the limit $M\to\infty$. The potential (\ref{3d_pot}) can be written in a more compact way
 as
\begin{equation}
\label{3d_pot2}
m(x_1,x_2,x_3)=m+(M-m)\prod_{i=1}^3[\theta(x_i-L_i)+\theta(-x_i)]
\end{equation}
where $\theta(x)$ is the step function.
For the solutions inside $V$ we expect to have combinations of free Dirac spinors with positive energy,
which have the general form
%
%{
%
\be
\psi = N
e^{i \vec k \cdot \vec r}
\left( \begin{array}{c}
\chi \\
{} \\
r \hat k \cdot \vec \sigma \chi \end{array} \right)
\la{3ddiracsoln}
\ee
where $\vec k$ is the wave vector and $r=\hbar |\vec k| c/(E+mc^2)$.

In the non-relativistic case (i.e., for the Schr\"odinger equation
with free particles confined in the volume $V$), since the
potential (\ref{3d_pot2}) is separable, one can use the product
ansatz for the wave function \be \psi = N\prod_{j=1}^3
\big(B_j~e^{ik_j x_j}+C_j~e^{-ik_j x_j}\big) \label{prod_ans} \ee
and thus turn the three-dimensional problem into a set of three independent
one-dimensional problems,
the total energy being just the sum of the one-dimensional energies.
In the relativistic case, however, one cannot use such
an ansatz because of the spinor structure of the wavefunctions.
Nevertheless, we may use that feature of the non-relativistic wavefunction
as a guide to find the right combination of spinors (\ref{3ddiracsoln}).
Indeed, we may require that in the non-relativistic limit,
when the lower component of the spinor vanishes,
the space part of the remaining 2-component spinor
be of the form (\ref{prod_ans}).
A wavefunction that meets this requirement is
\bea
\psi =\hspace{-1ex}
\left( \begin{array}{c}
\le[
\prod_{j=1}^3
\le(
B_j e^{i k_j x_j} + C_j e^{-ik_j x_j} \ri)
\ri] \chi
\\
{} \\
\hspace{-1ex}
\sum_{m=1}^3
\le[
\prod_{j=1}^3
\le( B_j e^{i k_j x_j} + C_j (-1)^{\d_{jm}} e^{-ik_j x_j} \ri)
r \hat k_m
 \ri]
\sigma_m \chi \hspace{-2ex}
\end{array}
\right)\,\,
\la{psi3d}
\eea
where an overall normalization has been set to unity. It can be easily shown that the above
is a superposition, with appropriate products of the
coefficients $B_i$ and $C_i$, of the following 8 eigenfunctions, for all possible combinations
 of $\epsilon_i~(i=1,2,3)$, with $\epsilon_i = \pm 1$
\bea
\Psi &=& e^{i\sum_{i=1}^3 \epsilon_i k_i x_i}
\left( \begin{array}{c}
\chi \\
{} \\
r \sum_{i=1}^3 \epsilon_i \hat k_i \sigma_i \chi
\end{array} \right) \ .
\eea
Each one of these eigenfunctions is of type (\ref{3ddiracsoln}), and they
represent plane waves travelling in the 8 directions $(\pm k_1,\pm k_2,\pm k_3)$
all with the same momentum magnitude $p=\hbar k=\hbar\sqrt{k_1^2+k_2^2+k_3^3}$.

Again, we impose the MIT bag boundary conditions
$\pm i \b \a_l \psi = \psi~,l=1,2,3$, with the $+$ and $-$ signs
corresponding to $x_l=0$ and $x_l=L_l$ respectively, ensuring vanishing
flux through all six boundaries. First, we write the above
boundary condition for any $x_l$, for the wavefunction given in Eq.(\ref{psi3d}).
This yields the following $2$-component equation
\bea
\hspace{-2ex}
&\pm&
\hspace{-1ex}
\left( \begin{array}{c}
i
\sum_{m=1}^3
\le[
\prod_{j=1}^3
\le(
B_j e^{i k_j x_j} + C_j(-1)^{\d_{jm}} e^{-ik_j x_j}\ri) r\s_l~ \hat k_m \s_m
\ri] \chi
\\
{}\\
-i \le[ \prod_{j=1}^3
\le(
B_j e^{i k_j x_j} + C_j e^{-ik_j x_j}
\ri)
\ri] \s_l \chi
\end{array}
\right)\hspace{-4ex} \nn \\
&=& \psi ~.
\label{3dbc}
%\left( \begin{array}{c}
%\end{array}
%\right)
\eea
By equating the upper components of the spinors
on each side of this equation and invoke the arbitrariness of $\chi$
one gets the matrix relation
\bea
&& \prod_{j=1}^3
\le(
B_j e^{ik_j x_j} + C_j e^{-ik_j x_j }
\ri)I\nn\\
&& =
\pm
 i \sum_{m=1}^3
\Big[ \prod_{j=1}^3
\le(
B_j e^{i k_j x_j} + C_j(-1)^{\d_{jm}} e^{-ik_j x_j}
\ri) r \hat k_m \s_l \s_m   \Big ]~.
\la{row1}
\eea
In the same fashion,
equating the lower components of the spinors
on each side of Eq.(\ref{3dbc}) and right multiply the resulting matrix equation
by $\pm i\,\s_l$ one gets
\bea
&& \pm i\sum_{m=1}^3
\Big[
\prod_{j=1}^3
\Big( B_j e^{i k_j x_j} + C_j (-1)^{\d_{jm}} e^{-ik_j x_j} \Big)
r \hat k_m
 \sigma_m \s_l\Big]
\nn \\
&& =
\prod_{j=1}^3
\le(
B_j e^{i k_j x_j} + C_j e^{-ik_j x_j}
\ri)I~.
\la{row2}
\eea
\par\noindent
Adding these two equations one gets
\bea
&&2\prod_{j=1}^3
\le(
B_j e^{i k_j x_j} + C_j e^{-ik_j x_j}
\ri)I\nn\\
&&=\pm i\sum_{m=1}^3
\Big[
\prod_{j=1}^3
\le( B_j e^{i k_j x_j} + C_j (-1)^{\d_{jm}} e^{-ik_j x_j} \ri)
r \hat k_m
 \{\s_l,\sigma_m\}\Big]\nn\\
&&=\pm 2i
\Big[
\prod_{j=1}^3
\le( B_j e^{i k_j x_j} + C_j (-1)^{\d_{jl}} e^{-ik_j x_j} \ri)
r \hat k_l
\Big] I
\eea
where the relation $\{\s_l,\sigma_m\}=2\delta_{lm}I$ was used. Noticing that for $j\not=l$
the terms in the products are the same in both sides of the equation,
one can divide both sides by
$\prod_{j\not=l}^3
\le( B_j e^{i k_j x_j} + C_j  e^{-ik_j x_j} \ri)$
and obtain, for each $l$,
\be
B_l e^{i k_l x_l} + C_l e^{-ik_l x_l}=\pm i
\le( B_l e^{i k_l x_l} - C_l e^{-ik_l x_l} \ri)
r \hat k_l \ .
\la{bc5}
\ee
Note that the boundary condition (\ref{3dbc}) gave rise to
3 independent condition pairs, one for each spatial dimension.
%We shall see
%momentarily that the $x_i~(i\neq k)$ dependence of the $F$-terms (via $f_{\bar k}$)
%will eventually drop out as well.
Eq.(\ref{bc5}) yields, at $x_l=0$ and $x_l=L_l$ respectively
\be
B_l  + C_l =i
\le( B_l - C_l \ri) r \hat k_l
\la{bc7}
\ee
and
\be
B_l e^{i k_l L_l} + C_l e^{-ik_l L_l}=- i
\le( B_l e^{i k_l L_l} - C_l e^{-ik_l L_l} \ri)
r \hat k_l ~.
\la{bc8}
\ee
Comparing Eqs.(\ref{bc7}) and (\ref{bc8}) with Eqs.(\ref{cond1}) and (\ref{cond2}),
we see that we can write, in a similar way as in the previous section,
\be
\frac{i r \hat k_l -1}{ir \hat k_l +1}=e^{ik_lL_l}
\ee
and thus
\be
\tan k_l L_l = \frac{2r \hat k_l}{r^2 \hat k_l^2-1}=
\frac{2(E+mc^2)\hbar k_l}{\hbar^2(k_l^2-k^2)c-2mc(E+mc^2)}~.
\la{3dquant3}
\ee
These are a set of three coupled transcendental equations for $k_1,k_2,k_3$
that need to be solved to find the energy eigenvalues. Again, the solutions with
$k_l=0$ are excluded.
In the non-relativistic limit, $E\sim m c^2,\hbar k_l/(mc)=\epsilon_l\ll 1$ so that
\be
k_l L_l \sim \arctan \le(- \epsilon_l\ri)\sim n_l\pi
\la{3dquantnr}
\ee
where $n_l=1,\ldots$ so that we recover the well-known quantization conditions
for a non-relativistic fermion gas confined in a cubic box.

On the other hand, in the ultra-relativistic limit, when $E\sim \hbar k c\gg mc^2$
and $\hbar k_l/(mc)\gg 1$, one gets
\be
k_l L_l \sim \arctan \le(- \alpha\frac{\hbar k_l}{mc}\ri)\sim (n_l-1/2)\pi
\la{3dquantur}
\ee
where $\alpha$ is a coefficient whose value depends on relative magnitude between $k_l$ and the other
components (i.e., $\alpha=1$ if $k_l\gg k_{l'},~\,l'\not=l$).

The size of the box plays a fundamental role in determining the relativistic behavior
of the solutions. Indeed, one may write
\be
k_l L_l=\frac{\hbar k_l}{mc}\,\frac{L_l}{L_C}
\la{k-l}
\ee
where $L_C=\hbar/(mc)$ is the Compton
wavelength. Given that the righthand side of Eq.(\ref{3dquant3}) is always negative (we take every $k_l$ to be positive) and therefore
its first solution such that
$\pi/2<k_l L_l< \pi$, we see that when $L_l\gg L_C$ and $L_l\ll L_C$ we get the non-relativistic and ultra-relativistic limits,
respectively. To illustrate this point, we write the eigenvalue equation in terms
of ratio $L_C/L_l$. When $k_l$ is much bigger than the other components, Eq.(\ref{3dquant3}) reads
\be
\tan x_l = -x_l\, \frac{L_C}{L_l}
\la{3dquant3_2}
\ee
where $x_l=k_lL_l$. The graphical solution of this equation is shown in Fig. 1.

\vskip.5cm
\begin{figure}[!ht]
\begin{center}
\includegraphics[width=10cm]{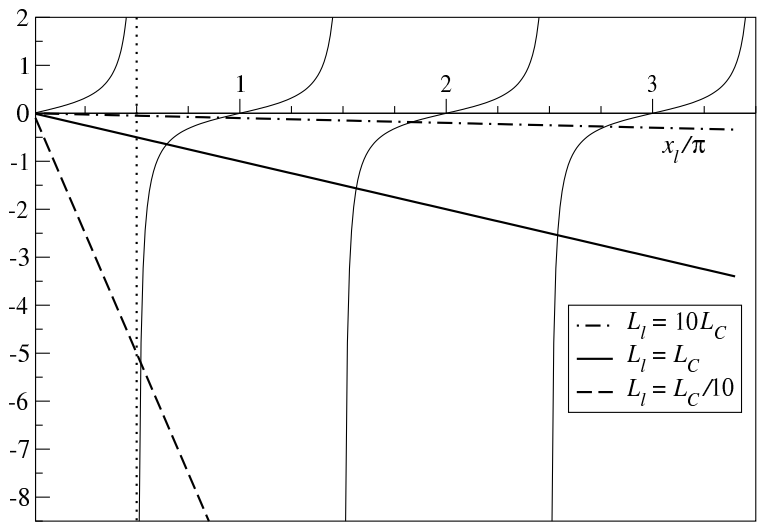}
\end{center}
\caption{Graphical solution of equation (\ref{3dquant3_2})
for three values of the ratio $L_l/L_C$. The dotted vertical corresponds to
$x_l=\pi/2$.}
\label{Fig1}
\end{figure}

From Fig. 1, one can check that indeed the size of box determines the relativistic nature of the
solutions and also that the first non-relativistic solution is
$k_l \sim \pi/L_l$ and
the first ultra-relativistic solution is $k_l\sim \pi/(2 L_l)$. Of course, the degree of relativity increases
for the higher energy solutions. Note that if the condition of $k_l$ being much bigger than the other components
is relaxed, this would amount to a small change in the coefficient of $x_l$ in the right-end side of
Eq.(\ref{3dquant3_2}), provided that $k_l=\max_i(k_i)$, so that the previous conclusions would still hold.
Note that all these conclusions assume that one has always a non-zero mass for the fermions.

In Fig.~2, it is depicted the energy spectra corresponding to the solutions of Eqs.~(\ref{3dquant3})
for $L_l=L_C/10$, $L_l=L_C$, $L_l=10 L_C$, considering a cubic box $L_1=L_2=L_3$. For convenience,
the values plotted are of the logarithm of the scaled kinetic energy $E/(mc^2)-1$. The levels plotted
correspond to the first 27 levels for each value of $L_l$. However, due to the symmetry of equations
(\ref{3dquant3}) with respect to the interchange of the $k_l$'s among them,
some of the levels are degenerate (besides, of course, the spin degeneracy). If $k_1\not=k_2\not=k_3$,
we have a $3!=6\,$-fold degeneracy, while when $k_l\not=k_{l'}=k_{l''}\ , \{l,l',l''\}=\hbox{permutations of }\{1,2,3\}$,
one has a 3-fold degeneracy. This reduces the 27 levels to 10 levels of distinct energies.

\vskip.2cm

\begin{figure}[!ht]
\begin{center}
\includegraphics[width=10cm]{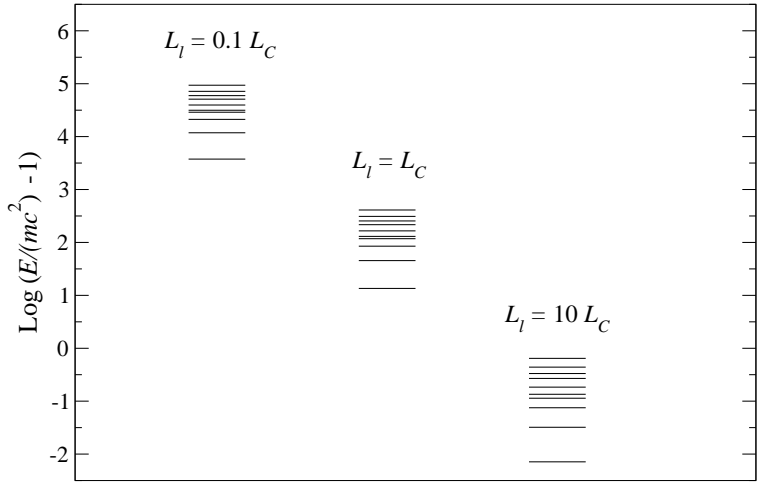}
\end{center}
\caption{Scaled kinetic energy spectrum of the first 27 solutions of equation (\ref{3dquant3})
for three values of the ratio $L_l/L_C$.}
\label{Fig2}
\end{figure}

This degeneracy is completely analogous to the one found in the non-relativistic case, in which
the kinetic energy is given by $E-mc^2=h^2\pi^2/(2mL^2)(n_1^2+n_2^2+n_3^2)\ ,n_1,n_2,n_3=1,\ldots$, $L$ being the
size of the cubic box.
In fact, one can use the non-relativistic quantum numbers $n_l$
to classify the quantum states in the present case. This is presented in Table \ref{Table1},
which contains the quantum numbers, the level degeneracy, the values of $k_lL_l/\pi$ and
corresponding energy values for the first  6 distinct energy states.

\begin{table}[!t]
\renewcommand{\arraystretch}{1.3}
\begin{tabular}{|c|c|c|c|c|}
  \hline
  \hline
  \rule[3.5ex]{0pt}{0pt}\ $(n_1,n_2,n_3)$&\ degen.\ & $(k_1,k_2,k_3) (\times\, L_l/\pi)$&\ $L_l/L_C$\ & \ $E/(mc^2)$\ \\[1ex]
  \hline
         &   &\ (0.674129, 0.674129, 0.674129)\ &  0.1  &  36.6957  \\
(1,1,1)  & 1 & (0.730735, 0.730735, 0.730735) &    1  &  4.10004  \\
         &   & (0.914156, 0.914156, 0.914156) &   10  &  1.11689  \\
\hline
        &   & (0.761157, 0.761157, 1.5664) &    0.1  &  59.718  \\
(1,1,2) & 3 & (0.789821, 0.789821, 1.61153) &     1  &  6.24063  \\
        &   & (0.917935, 0.917935, 1.8383) &     10  &  1.22469  \\
\hline
        &   & (0.800534, 1.62894, 1.62894) &    0.1  &  76.6236  \\
(1,2,2) & 3 & (0.820262, 1.66176, 1.66176) &      1  &  7.88349  \\
        &   & (0.921162, 1.84449, 1.84449) &     10  &  1.32488  \\
\hline
        &   & (0.819801, 0.819801, 2.53383) &   0.1  &  87.5453  \\
(1,1,3) & 3 & (0.835499, 0.835499, 2.56592) &     1  &  8.93086  \\
        &   & (0.923098, 0.923098, 2.77709) &    10  &  1.38902  \\
\hline
        &   & (1.66969, 1.66969, 1.66969) &     0.1  &  90.8601  \\
(2,2,2) & 1 & (1.69565, 1.69565, 1.69565) &       1  &  9.28075  \\
        &   & (1.84989, 1.84989, 1.84989) &      10  &  1.41889  \\
\hline
        &   & (0.838015, 1.69214, 2.57123) &    0.1  &  100.225  \\
(1,2,3) & 6 & (0.850724, 1.71438, 2.59869) &      1  &  10.1883  \\
        &   & (0.92567, 1.85317, 2.7841) &       10  &  1.47937  \\
%(2,2,3) & 3 & (1.71607,1.71607,2.6008) &      0.1  &  111.759  \\
%(2,2,3) & 3 & (1.73505,1.73505,2.62479) &       1  &  11.3323  \\
%(2,2,3) & 3 & (1.85759,1.85759,2.79036) &      10  &  1.56512  \\
%(1,3,3) & 3 & (0.859576,2.6185,2.6185) &      0.1  &  119.434  \\
%(1,3,3) & 3 & (0.869053,2.64033,2.64033) &      1  &  12.0857  \\
%(1,3,3) & 3 & (0.929364,2.79423,2.79423) &     10  &  1.62063  \\
%(2,3,3) & 3 & (1.74658,2.63958,2.63958) &     0.1  &  129.479  \\
%(2,3,3) & 3 & (1.76165,2.65918,2.65918) &       1  &  13.0847  \\
%(2,3,3) & 3 & (1.86403,2.79953,2.79953) &      10  &  1.69999  \\
%(3,3,3) & 1 & (2.66856,2.66856,2.66856) &     0.1  &  145.211  \\
%(3,3,3) & 1 & (2.68519,2.68519,2.68519) &       1  &  14.6454  \\
%(3,3,3) & 1 & (2.8074,2.8074,2.8074) &         10  &  1.82582  \\
  \hline
  \hline
\end{tabular}
\caption{Non-relativistic quantum numbers, degeneracies, values of $k_l$ in units of $\pi/L_l$ and scaled energies
 for the first 6 distinct energy solutions of equation (\ref{3dquant3}). For each level are presented the values for
  $L_l/L_C=0.1, 1, 10$.}
\label{Table1}
\end{table}

One can check from the values presented in the table that indeed, for small values of $L_l/L_C$,
$k_l$ approaches $(n_l-1/2) \pi/L_l$ while for higher values they approach $n_l \pi/L_l$, the non-relativistic
value, as suggested by Fig.~1.

Finally, we will make some considerations regarding density of states of relativistic fermion gases. From Fig.~1 and
Table \ref{Table1} one sees that as $L_l$ gets bigger each allowed value for $k_l$ is separated by $\pi/L_l$ as in the non-relativistic fermion gas. For fermion gases confined in large volumes with many particles,
as considered for instance in Statistical Physics, the Dirac gas in a 3-D box behaves as a non-relativistic gas.
For sizes $V\sim L_C^3$, for which the fermions are relativistic, the density, measured by the separation of $k_l$
values in units of $\pi/L_l$, tends to increase, as can be seen from Fig.~1 and Table \ref{Table1}, since each solution is separated by less than $\pi/L_l$. For smaller volumes, we checked that we recover again the non-relativistic density of states (NRDS) when the quantum numbers changing are relatively high.
In general, we can say that, for higher energy levels, the density of states tends to be the same as the NRDS. On the other hand, for small quantum numbers, in relativistic (and ultra-relativistic) conditions, we get higher densities than the NRDS.

%%%%%%%%%%%%%%%%%%%%%%%%%%%%%%
\section{Conclusions}
%%%%%%%%%%%%%%%%%%%%%%%%%%%%%%
%
%
In this paper, we have solved the problem of a relativistic spin-1/2 particle in a three-dimensional
square box using the Dirac equation. We studied both the non-relativistic and ultra-relativistic limits and
showed that these are related to the size of the box. The scale for gauging the relativity of
the solution is the Compton wavelength, such that free fermions confined in boxes with sizes many times the Compton
wavelength behave as non-relativistic particles. We solved the three coupled transcendental equations which give the
energy eigenvalues of the Dirac particle in a 3D box and used the non-relativistic quantum numbers of quantum particle
in a box to classify the corresponding states. We also showed that the density of states is in general
higher than the non-relativistic one, but for higher energy states it will tend to be equal to
non-relativistic density of states, in which the allowed values for the wavenumber components $k_l,\,l=1,2,3$
are separated by $\pi/L_l$, where $L_l$ is the respective box dimension.
%

%%%%%%%%%%%%%%%%%%%%%%%%%%%%%%%%%%%%%%%%%%%%%%%%%%%%%%
\vskip1cm
\noindent\textbf{Acknowledgments}
\vskip.3cm

This work was supported in part by the Natural
Sciences and Engineering Research Council of Canada and by the
Perimeter Institute for Theoretical Physics, as well as the projects
PTDC/FIS/64707/2006 and CERN/FP/109316/2009.

%%%%%%%%%%%%%%%%%%%%%%%%%%%%%%%%%%%%%%%%%%%%%%%%%%%%%

%\section*{References}

\end{document}